\documentclass{aa}

\usepackage[dvips]{graphicx}
\usepackage{amsmath} 
\usepackage{dcolumn}
\usepackage[figuresright]{rotating}    

\newcommand{\widthni}{{\sc width9}}
\newcommand{\cf}{{cf.}}
\newcommand{\vs}{{vs.}}
\newcommand{\last}{{LLB}}
\newcommand{\str}{{Str\"omgren}}
\newcommand{\balmernine}{{\sc balmer9}}
\newcommand{\simbad}{{\sc simbad}}
\newcommand{\romer}{{\sc mons/r\o mer}}
\newcommand{\corot}{{\sc corot}}

\newcommand{\elodie}{{\sc elodie}}
\newcommand{\inter}{{\sc inter-tacos}}
\newcommand{\gaudi}{{\sc gaudi}}

\newcommand{\sect}{Sect.}
\newcommand{\sects}{Sects.}

\newcommand{\atlasnine}{{\sc atlas9}}   

\newcommand{\topp}{{\em top}}
\newcommand{\bott}{{\em bottom}}

\newcommand{\feh}{[Fe/H]}

\newcommand{\templogg}{{\sc templogg}}                      
\newcommand{\vald}{{\sc vald}}                              
\newcommand{\hipparcos}{{\sc hipparcos}}                              
\newcommand{\hipp}{{\sc hipparcos}}                              
\newcommand{\teff}{\ensuremath{T_{\rm eff}}}                    
\newcommand{\logg}{\ensuremath{\log g}}                     
\newcommand{\loggf}{\ensuremath{\log gf}}                     
\newcommand{\exita}{\ensuremath{E_{\rm low}}}                     
\newcommand{\vsini}{\ensuremath{v\sin i}}                   

\newcommand{\fehphot}{[M/H]$_{\rm phot}$}
\newcommand{\halpha}{H\,$\alpha$} 
\newcommand{\kms}{\ensuremath{\text{km\,s}^{-1}}}           

\newcommand{\etal}{et~al.}

 \newcommand{\vwa}{{\sc vwa}}         

 \newcommand{\ie}{i.e.} 
 \newcommand{\eg}{e.g.} 
  \newcommand{\etc}{etc} 
\newcommand{\rms}{{\em rms}} 

\def\iet{\,{\sc i}} \def\ii{\,{\sc ii}}  
\def\ito{\,{\sc ii}}

\newlength{\dplotwidth}
\begin{document}

 \title{Abundance analysis of targets for 
          the COROT / MONS asteroseismology missions}

   \subtitle{II. Abundance analysis of the COROT main targets\thanks{Based on
           observations obtained with the 193~cm telescope 
           at Observatoire de Haute Provence, France}}

   \author{H.~Bruntt\inst{1}
        \and
         I.~F.~Bikmaev\inst{2}
        \and
         C.~Catala\inst{3} 
        \and
         E.~Solano\inst{4}
        \and 
         M.~Gillon\inst{5}
        \and
         P.~Magain\inst{5}
        \and 
         C.~Van't Veer-Menneret\inst{6}
        \and
         Ch.~St\"utz\inst{7}
        \and
         W.~W.~Weiss\inst{7}
        \and 
         D.~Ballereau\inst{6}
        \and 
         J.C.~Bouret\inst{8}
        \and 
         S.~Charpinet\inst{9}
        \and 
         T.~Hua\inst{8}
        \and 
         D.~Katz\inst{6}
        \and 
         F.~Ligni\`eres\inst{9}
        \and 
         T.~Lueftinger\inst{7}
          }
   \offprints{H.\ Bruntt, \email{bruntt@phys.au.dk}}
   \institute{Department of Physics and Astronomy, Aarhus University,
              Bygn.~520, DK-8000 Aarhus C, Denmark
        \and
            Department of Astronomy, Kazan State University,
            Kremlevskaya 18, 420008 Kazan, Russia        
        \and 
            Observatoire de Paris, LESIA, France  
        \and 
            Laboratorio de Astrof\'\i{}sica Espacial y F\'\i{}sica Fundamental, 
            P.\,O.\ Box 50727, 28080 Madrid, Spain
        \and
            Institut d'Astrophysique et de G\'eophysique, 
            Universit\'e de Li\`ege,     
            All\'ee du 6 Ao\^ut, 17, 
            4000 Li\`ege,            
            Belgium
        \and
            Observatoire de Paris, GEPI, France 
        \and
            Institut f\"ur Astronomie, Universit\"at Wien,
            T\"urkenschanzstrasse 17, A-1180 Wien, Austria 
        \and
            Laboratoire d'Astrophysique de Marseille, France
        \and 
            Laboratoire d'Astrophysique de l'OMP, CNRS UMR 5572,
            Observatoire Midi-Pyr\'en\'ees, 14, avenue Edouard Belin, F-31400
            Toulouse, France 
}


\date{Received ..... / Accepted .....}

\abstract{One of the goals of the ground-based 
support program for the \corot\ and \romer\ satellite missions
is to characterize suitable target stars for the
part of the missions dedicated to asteroseismology.
We present the detailed abundance analysis of nine of the 
potential \corot\ main targets using the semi-automatic software \vwa.
For two additional \corot\ targets we could not perform the analysis 
due to the high rotational velocity of these stars. 
For five stars with low rotational
velocity we have also performed abundance analysis by a classical
equivalent width method in order to test the reliability of the
\vwa\ software. The agreement between the different methods is good.
We find that it is necessary to measure abundances extracted from each line 
relative to the abundances found from a spectrum of the Sun in order to remove
systematic errors. 
We have constrained the global atmospheric parameters \teff, \logg, and 
[Fe/H] to within $70-100$~K, $0.1-0.2$~dex, and 0.1~dex for five stars 
which are slow rotators ($\vsini < 15$~\kms). 
For most of the stars stars we find good agreement with the parameters found 
from line depth ratios, \halpha\ lines, Str\"omgren indices, previous spectroscopic studies, 
and also \logg\ determined from the \hipparcos\ parallaxes. For the fast
rotators ($\vsini > 60$~\kms) it is not possible to constrain the atmospheric
parameters.
   \keywords{ 
    stars: abundances -- 
    stars: fundamental parameters -- }
} 
\maketitle


\section{Introduction}
\corot\ (COnvection, ROtation, and planetary Transits) is a small space mission, dedicated to 
asteroseismology and the search for exo-planets (Baglin \etal~2001). Among the targets of the asteroseismology 
part of the mission, a few bright stars will be monitored continuously over a period of 150 
days. These bright targets will be chosen from a list of a dozen F \& G-type stars, located in the 
continuous viewing zone of the instrument. The final choice of targets needs to be made early 
in the project, as it will impact on some technical aspects of the mission. A precise and reliable
knowledge of the candidate targets is required, in order to optimize this final choice. 

Among the information which needs to be gathered on the candidate targets, fundamental parameters
like effective temperature, surface gravity, and metallicity will play a major role in the selection
of targets for \corot. Projected rotation velocities, as well as detailed abundances of the main 
chemical elements will also be taken into account in the selection process. 
{\ bf Thus, the aim of this study is to obtain improved values for the fundamental parameters and
abundances of individual elements for the \corot\ main targets.}

This information on the targets will subsequently be used for the selected stars, in conjunction
with asteroseismological data obtained by \corot, to provide additional constraints for the 
modelling of the interior and the atmospheres of these stars.


In \sect~\ref{observations} we summarize the spectroscopic observations
and data reduction, in \sect~\ref{sec:fundamental} we discuss the
determination of the fundamental parameters from spectroscopy, 
photometry and \hipparcos\ parallaxes and we summarize previous 
spectroscopic studies of the target stars.
In \sect~\ref{sec:methods} we describe the three different 
methods we have used for abundance analysis. 
In \sect~\ref{constrain} we discuss how we constrain the
fundamental atmospheric parameters using abundance results for
a grid of models. In \sect~\ref{sec:abundances} we discuss the
abundances we have determined.
Lastly, we give our conclusions in \sect~\ref{conclusions}.

\begin{table} \centering
\caption[]{Log of the observations for the spectra of the proposed
\corot\ targets we have analysed. The signal-to-noise ratio in the
last column is calculated around 6\,500 \AA\ in bins of 3~\kms.
\label{tab:log}}
\begin{tabular}{r|rc|r|r}
\hline
HD      & \multicolumn{1}{c}{Date}           & UT start      & $t_{\rm exp}$   & S/N  \\
\hline		      		     			  
 43318  & 19-Jan-98      & 22:38         &  900            & 120 \\
 43587  & 14-Jan-98      & 22:32         & 1800            & 250 \\
 45067  & 15-Jan-98      & 22:20         & 1200            & 260 \\
 49434  & 17-Jan-98      & 22:33         & 600             & 160 \\
 49933  & 21-Jan-98      & 22:17         & 1200            & 210 \\
 55057  & 15-Jan-98      & 23:10         & 900             & 270 \\
 57006  & 10-Dec-00      & 00:41         & 1800            & 250 \\
171834  & 05-Sep-98      & 19:00         & 1800            & 300 \\
184663  & 18-Jun-00      & 01:30         & 1000            & 170 \\
 46304  & 17-Jan-98      & 22:05         & 600             & 170 \\
174866  & 04-Jul-01      & 00:39         & 1500            & 190 \\
\hline
\end{tabular}
\end{table}

\begin{table} \centering 
  \caption{Str\"omgren photometric indices of the \corot\
main targets taken from Hauck \& Mermilliod (1998).
HD~46304 and HD~174866 are shown separately: abundance
analysis has not been made for these two stars due to their high
\vsini.\label{tab:stromgren_corot}}
  
  \begin{tabular}{r|ccccc}
 \hline 
\multicolumn{1}{c|}{HD} & $V$  &  $b-y$ & $m_1$ &   $c_1$ & ${\rm H}_\beta$ \\
 \hline
 43318  & 5.65 & 0.322 & 0.154 &   0.446 &   2.644 \\
 43587  & 5.70 & 0.384 & 0.187 &   0.349 &   2.601 \\
 45067  & 5.87 & 0.361 & 0.168 &   0.396 &   2.611 \\
 49434  & 5.74 & 0.178 & 0.178 &   0.717 &   2.755 \\
 49933  & 5.76 & 0.270 & 0.127 &   0.460 &   2.662 \\
 55057  & 5.45 & 0.185 & 0.184 &   0.876 &   2.757 \\
 57006  & 5.91 & 0.340 & 0.168 &   0.472 &   2.625 \\
171384  & 5.45 & 0.254 & 0.145 &   0.560 &   2.682 \\
184663  & 6.37 & 0.275 & 0.149 &   0.476 &   2.665 \\
\hline
 46304  & 5.60 & 0.158 & 0.175 &   0.816 &   2.767 \\
174866  & 6.33 & 0.122 & 0.178 &   0.960 &   2.822 \\
\hline
  \end{tabular}
\end{table}

\section{Spectroscopic observations\label{observations}}

We have obtained spectra of each one of the 11 candidate main targets of \corot, 
using the \elodie\ spectrograph attached to 
the 1.93\,m telescope at Observatoire de Haute-Provence (OHP). \elodie\ is a fiber-fed 
cross-dispersed Echelle spectrograph, providing a complete spectral coverage of the 
3800--6800~\AA\ region, at a resolution of $R = 45\,000$ (Baranne \etal\ 1996). 
Table~\ref{tab:log} gives the log of the 
observations for the spectra used in this analysis.

\subsection{Data reduction\label{datareduction}}

We used the on-line \inter\ reduction package available at OHP 
(Baranne \etal\ 1996). 
This software performs bias and background light subtraction, spectral order 
localization, and finally extracts spectral orders using the optimal extraction procedure 
(Horne 1986). The high spatial frequency instrumental response is corrected by dividing the stellar 
spectra by the spectrum of a flat-field lamp. Wavelength calibration is provided by spectra of 
a Th/Ar lamp, using a two-dimensional Chebychev 
polynomial fitting to the centroid locations of the Th/Ar lines.

Special care was taken to correct for the grating blaze function, as differences in the 
spectrograph illumination between stellar and flat-field light beams usually result in an 
imperfect correction. Instead of using a flat-field spectrum, we have therefore used a high 
signal-to-noise spectrum of an O-type star, 10~Lac, to determine the blaze function. The 
orders of the 10~Lac spectrum were examined one by one, and the line-free regions of each 
order were identified. The blaze function at each spectrograph order was then represented by 
cubic splines fitted to these line-free regions, and the stellar spectra were 
subsequently divided by this newly determined blaze function. This procedure results in an 
adequate blaze correction, producing in particular a good match of adjacent orders in the 
overlapping regions. Note also that \elodie\ is a very stable instrument, so that only one 
spectrum of 10~Lac was used to define the blaze function, although the observations reported 
here span several years.


Since the overall blaze function was removed by a star of
much earlier spectral class than 
the observed F and G-type stars the continuum level was not
entirely flat. Hence we made cubic spline fits to make the final
continuum estimate. For the overlapping part of the orders 
we made sure that the overlap was better than 0.5\%. 

We are aware of the problem of fitting low-order splines to
correct the continuum. In this process we manually mark the points
in the spectrum which we assume are at the continuum level. 
When using the 
\vwa\ software (Bruntt \etal\ 2002, cf.\ \sect~\ref{vwa_desc})
we can inspect the fitted lines 
and in this way detect problems with the continuum level. 
In these cases we reject the abundances found for these lines
(\cf\ \sect~\ref{sec_comp_meth}).
Alternatively, one could select ``continuum windows'' from a 
synthetic spectrum or a spectrum of the Sun and use
this to correct the continuum level. This has been attempted
for one of our program stars in Sect.~\ref{sec:magain}.


\begin{table*} \centering \footnotesize
  \caption{Overview of the parameters of the proposed \corot\ main targets.
The first column is the HD number and column 2 is \teff\ determined from
line depth ratios with formal errors in parenthesis (Kovtyukh \etal~2003).
Column 3 gives the temperatures found from the \halpha\ wings (the internal error is 50~K).
Column 4--6 are the atmospheric parameters derived from Str\"omgren photometry
when using the \templogg\ software (typical errors are 200~K, 0.3~dex, and 0.2 dex).
Column 7 and 8 are the masses found from evolution tracks (cf.~Fig.~\ref{fig:mass}) and
$\log g_\pi$ values found from the \hipparcos\ parallaxes; 
the numbers in parenthesis are the estimated standard errors.
In column 9 we list \vsini\ where the typical error is 5--10\%.
\label{tab:fundamental}}
\setlength{\tabcolsep}{3pt} 
  \begin{tabular}{r|l|l|ccc|cc|r}
  \hline 
      & \multicolumn{1}{c|}{Line depth} & \halpha & \multicolumn{3}{c|}{Str\"omgren} & \multicolumn{2}{c|}{Evolution Tracks \&} & Spectral \\
      & \multicolumn{1}{c|}{ratio}      & wings       & \multicolumn{3}{c|}{indices}     & \multicolumn{2}{c|}{Parallax}  & synthesis \\ \hline
  HD  & \teff\ [K] & \teff\ [K]  & \teff\ [K] & \logg\ & \fehphot\ & $M/M_\odot$ & $\log g_\pi$ & \vsini\ [km/s] \\
  \hline 
   Sun & 5770(5)  &  $-$  & 5778$^a$ & 4.44$^a$ & $+0.00^a$  & 1$^a$ & 4.44$^a$& 2$^b$   \\ 
 43318 & 6191(17) &  6100 & 6400 & 4.19 &$-0.15 $   & 1.23(17)  & 3.96(14)  &    8       \\ 
 43587 & 5923(8)  &  5850 & 5931 & 4.31 &$-0.11 $   & 1.02(20)  & 4.29(15)  &    2.5$^b$ \\ 
 45067 & 6067(6)  &  5900 & 6038 & 4.03 &$-0.22 $   & 1.08(17)  & 3.96(15)  & $<$7$^b$  \\ 
 49434 &   $-$    &  6950 & 7304 & 4.14 &$-0.01 $   & 1.55(14)  & 4.25(11)  &   84  \\ 
 49933 &   $-$    &  6400 & 6576 & 4.30 &$-0.45 $   & 1.17(18)  & 4.20(14)  &   14  \\ 
 55057 &   $-$    &  6750 & 7274 & 3.61 &$+0.10 $   & 2.12(22)  & 3.66(12)  &  120  \\ 
 57006 & 6181(7)  &  6000 & 6158 & 3.72 &$-0.13 $   & 1.28(17)  & 3.58(16)  &    9  \\ 
171834 &   $-$    &  6550 & 6716 & 4.03 &$-0.22 $   & 1.40(17)  & 4.13(13)  &   63  \\
184663 &   $-$    &  6450 & 6597 & 4.25 &$-0.17 $   & 1.29(14)  & 4.19(15)  &   53  \\
\hline	     	      								    
 46304 &   $-$    &  7050 & 7379 & 3.93 &$-0.09$    & 1.68(14)  & 4.18(11)  &  200  \\ 
174866 &   $-$    &  7200 & 7865 & 3.86 &$-0.18$    & 1.77(14)  & 3.86(15)  &  165  \\
\hline
\multicolumn{9}{l}{$^a$ The fundamental parameters for the Sun are also given in the table, although we have not}\\
\multicolumn{9}{l}{determined its parameters; the exception is our estimate of \teff\ from line depth ratios.}\\
\multicolumn{9}{l}{$^b$ With the resolution of the \elodie\ spectrograph ($R=45.000$) it is not possible to measure}\\
\multicolumn{9}{l}{\vsini\ below 7 \kms\ directly.}\\

\end{tabular}
\end{table*}


\section{Fundamental atmospheric parameters\label{sec:fundamental}}

In this Section we will discuss previous results for the
fundamental atmospheric parameters of the proposed
\corot\ targets. We will present the results from
the calibration of \str\ photometry,
line depth ratios, \halpha\ line wings,
and \hipparcos\ parallaxes. 

\subsection{Str\"omgren photometry\label{sec:stromgren}}

The \str\ indices of the target stars are taken from 
the catalogue of Hauck \& Mermilliod (1998) 
and are listed in Table~\ref{tab:stromgren_corot}.
We have used the software {\sc templogg} 
(Rogers~1995, see also Kupka \& Bruntt~2001) to
find the appropriate calibration to obtain the basic parameters, 
\ie\ $T_{\rm eff}, \log g$, and \fehphot.
These parameters are given in Table~\ref{tab:fundamental}. 
The accuracy of the parameters \teff, \logg, and 
\fehphot\ are around 200--250~K, 0.3~dex, and 0.2~dex 
according to Rogers (1995).

A catalogue of \str\ indices determined for all primary and
secondary targets for \corot\ is available from 
the \gaudi\ database\footnote{{\tt http://ines.laeff.esa.es/corot/}}.
We also used these data to get the fundamental parameters. 
For most stars both \teff\ and \logg\ agree within the 
uncertainties quoted above. But for HD~43587 and
HD~45067 we find a large discrepancy, \ie\ \logg\ higher by 
0.2/0.3 dex and a higher \teff\ by 350/300~K for the two
stars, respectively. This is a clear indication that it is
worthwhile to use several methods to try to determine the
fundamental parameters.

\subsection{Temperature Calibration from Line Depth Ratios\label{sec:kovtyukh}}

Kovtyukh \etal\ (2003) have used spectra of 181 main sequence F-K type stars
to calibrate the dependence of line depth ratios on \teff.
They used observed spectra from ELODIE (the same spectrograph we used)
and their sample of stars consisted of stars for
which \teff\ is well determined, \eg\ using the infrared flux method.

We measured line depths by fitting a Gaussian profile to
each line, \ie\ the depth of the Gaussian defines the line depth. 
We used \teff\ from the Str\"omgren photometry as our initial guess 
(assuming the error is $\sigma($\teff$)=250$~K) to select which of the 
calibrations by Kovtyukh \etal\ (2003) that were valid.
We note that for each pair of lines defining the ratio calibration 
the species of elements are typically different (\eg\ Si/Ti, Fe/S \etc.), 
but we refer to Kovtyukh \etal\ (2003) for details. 
Typically 50--80 line depth ratios between lines could be used.
We then calculated the mean \teff\ and
rejected 3\,$\sigma$ outliers and recalculated \teff. 
The calibrations are only valid in the range 4000--6150~K so only the  
four coolest \corot\ target stars could be used with this method. 

The results are given in Table~\ref{tab:fundamental} for these
stars and the Sun. The \teff\ we determine from the spectrum of the Sun
agrees with the canonical value of \teff$=5777$~K.
The quoted errors of 5-17~K on the temperatures are formal errors, and
systematic errors of the order 50-100~K must be added.

\begin{figure} 
\hskip -0.7cm
\includegraphics[width=9.2cm]{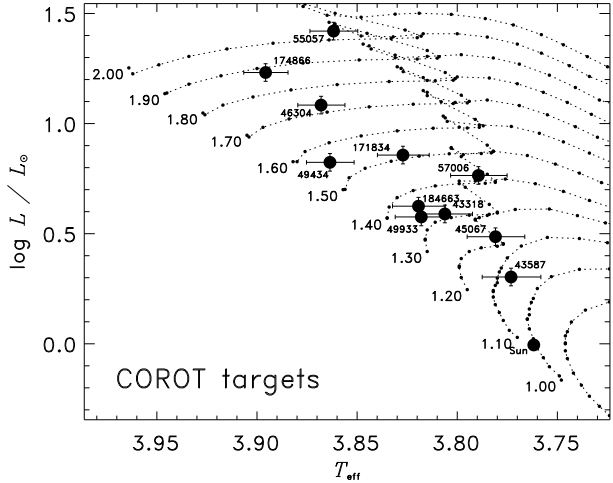} 
    \caption{Evolution tracks from Girardi \etal\ (2000) for solar metallicity.
The positions of the proposed \corot\ targets and the Sun are shown. 
\label{fig:mass}}
\end{figure}

\subsection{Temperatures from \halpha\ lines\label{sec:halpha}}

We have estimated \teff\ of the stars considered in
this study using their \halpha\ line profiles, following a method
proposed by Cayrel \etal~(1985) and Cayrel de Strobel \etal~(1994). 
The method is based on the property of \halpha\ that it is 
insensitive to any atmospheric parameter except \teff\ in
in the range between 5000--8500~K.

In order to overcome problems due to continuum placement, we compute
the ratio of the observed \halpha\ profile to that of the Sun, observed
with the same instrumental configuration (spectrum of the solar reflected
light on the moon surface), then compare it to the corresponding ratio of
theoretical profiles computed from a grid of models. The best fit between
computed and observed profile ratios gives the effective temperature with
an internal error bar of about $\pm50$~K.

We have used the new grids of \atlasnine\ models 
presented in Heiter \etal~(2002), choosing the models with MLT convection
treatment for \teff\ lower than 8750~K, with the low value for the efficiency
parameter of the convection, following Fuhrmann \etal\ (1993), 
Axer \etal\ (1994), and Heiter \etal\ (2002). 

The \halpha\ line profile is computed using 
\balmernine\ (Kurucz 1998) which includes the Vidal \etal\ (1973) unified
theory for the Stark Broadening and also takes into account 
the self resonance mechanism in the computation 
of the \halpha\ profile.

This method if found to be reliable for \teff\ between 5500 and 8500~K. 
Below 5500~K, the \halpha\ profile is not sensitive enough to \teff,
and does not constitute a good temperature tracer. Above 8500~K, the
\halpha\ wings depend only slightly on \teff, and start to become sensitive
to gravity.

We find that the final \teff\ estimate is insensitive 
to the choice of convection description. On the other hand 
Heiter \etal\ (2002) find that the $(b-y)$ index and hence
\teff\ found from the \str\ indices is sensitive to the 
convection description.

The \teff\ measurements with this method are also 
reasonably insensitive to rotation, 
for \vsini\ below 80--100 \kms.

\begin{figure} 
\centering
\includegraphics[width=8.1cm]{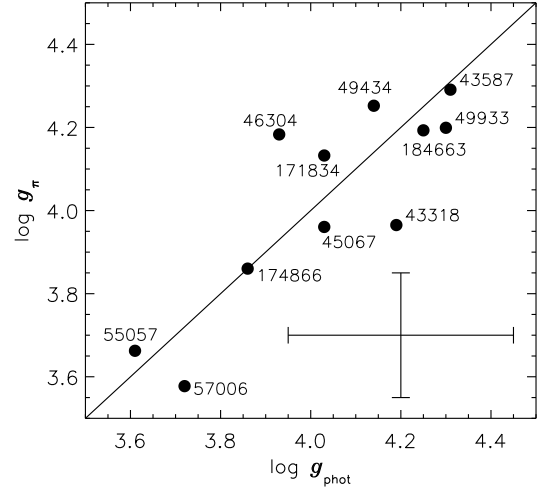} 
    \caption{Comparison of the $\log g_\pi$ values found from the
\hipparcos\ parallaxes and $\log g_{\rm phot}$ from \templogg\ using
Str\"omgren indices. The straight line is added to aid the eye.
Typical error bars are shown.
\label{fig:logg}}
\end{figure}

\subsection{Comparison of temperature estimators}

Except for HD~43318 there is very good agreement between \teff\ from 
the Str\"omgren calibration (\templogg) and the line depth ratios.
However, \teff\ of HD~43318 is on the border of the valid range of the
latter calibration. The \teff\ from the \halpha\ lines agrees with the 
other two methods but only for stars with low \vsini.
The stars in our sample with moderate \vsini\ also have the 
highest \teff, and the discrepant results for the \halpha\ lines
may be an indication that this method cannot be used for these stars.


\subsection{Using \hipparcos\ parallaxes to determine \logg\label{hipp}}

We have used \hipparcos\ parallaxes (ESA, 1997) to find the
surface gravities of the stars. We use the equation
$\log g_\pi = 4 [T_{\rm eff}] + [M] + 2 \log \pi + 0.4 (V + BC_V + 0.26) + 4.44$,
where $[T_{\rm eff}] = \log (T_{\rm eff}/T_{{\rm eff}\,\odot})$ and 
$[M] = \log (M/M_\odot)$. 
We use \teff\ from \templogg\ and we determine
the bolometric correction $BC_V$ from Bessell \etal\ (1998): we used
their results for solar metallicity and no overshoot.
We have estimated the masses of the stars by using the 
evolution tracks by Girardi \etal\ (2000).
The evolution tracks for solar metallicity is shown in Fig.~\ref{fig:mass}
in which the locations of the \corot\ targets are also shown along with
error bars.
We have made interpolations in models for metallicity 
$Z = 0.019$ (solar) and $Z = 0.008$. 
For example, HD~49933 has the lowest metallicity in the sample:
the derived mass is $M/M_\odot = 1.17$, significantly lower
than estimated from its position relative to the solar metallicity 
isochrones shown in Fig.~\ref{fig:mass}.

The determination of \logg\ when using the calibration of 
\str\ indices and the \hipparcos\ parallax are shown in Fig.~\ref{fig:logg}. 
Mean error bars which are representative for all the stars 
are shown in the lower right corner. 
Note that the two most important error sources 
for $\log g_\pi$ are \teff\ and the mass.
The results for $\log g_\pi$ and $\log g_{\rm phot}$ agree very well. 
The mean difference is $(\log g_{\rm phot} - \log g_\pi) = 
0.05 \pm 0.22$. 


\begin{table}
\caption{
Results of various efforts to determine the
parameters of the \corot\ targets. The results of this 
study are given as references 1a--c 
(corresponding to method A--C as described in Sect.~\ref{sec:methods}).
The references are 
given in Table~\ref{tab:study2}.\label{tab:study}}
\begin{tabular}{r|lllr}
HD      &  \teff     & \logg        & [Fe/H]        & Ref.  \\ \hline
 43318  &  6190(90)  &   3.70(0.15) & $ -0.21(11) $ &   1a  \\ 
     -  &  6347      &   4.07       & $ -0.17     $ &    4  \\ 
     -  &  6257      &   4.09       & $ -0.19     $ &    7  \\ 
     -  &  6320(100) &   4.5(0.2)   & $ +0.0(0.15)$ &   9a  \\
     -  &  6250(250) &   4.0(0.5)   & $ -0.3(0.2) $ &   9b  \\ \hline
 43587  &  5870(60)  &   4.20(0.15) & $ -0.09(0.11)$&   1a  \\ 
     -  &  5780(100) &   4.25(0.15) & $ -0.05(0.11$ &   1b  \\ 
     -  &  5867(20)  &   4.29(0.06) & $ -0.05(0.08)$&   1c  \\ 
     -  &  5775      &              & $ -0.08(0.04)$&    5  \\
     -  &  5931(100) &              & $ -0.00(0.05)$&    6  \\  
     -  &  5720(100) &   4.3(0.4)   & $ -0.2(0.10)$ &   9a  \\
     -  &  6000(250) &   4.5(0.5)   & $ -0.1(0.2) $ &   9b  \\
     -  &  5795      &              & $ -0.03     $ &   11  \\ 
     -  &            &              & $ -0.03     $ &   12  \\ \hline 
 45067  &  5970(100) &   3.80(0.15) & $ -0.17(0.11)$&   1a  \\ 
     -  &  6050(100) &   4.00(0.15) & $ -0.04(0.11)$&   1b  \\ 
     -  &  6056      &   4.17       & $ -0.16     $ &    2  \\ 
     -  &  5940(100) &   3.8(0.35)  & $ -0.1(0.08)$ &   9a  \\
     -  &  6000(250) &   4.0(0.5)   & $ -0.1(0.2) $ &   9b  \\  \hline
 49434  &  7300(200) &   4.4(0.4)   & $ -0.04(0.21)$&   1a  \\
     -  &            &              & $ -0.13(14) $ &    3  \\ 
     -  &  7240(150) &   4.0(0.25)  & $ -0.1(0.25)$ &   9a  \\
     -  &  7250(250) &              & $           $ &   9b  \\ \hline
 49933  &  6780(70)  &   4.3(0.2)   & $ -0.30(0.11)$&   1a  \\ 
     -  &  6472      &   4.17       & $ -0.35     $ &    2  \\ 
     -  &  6595      &   4.16       & $ -0.43     $ &    4  \\ 
     -  &  6300      &   4.5        & $ -0.88     $ &    8  \\ 
     -  &  6600(150) &   4.3(0.25)  & $ -0.6(0.15)$ &   9a  \\
     -  &  6500(250) &   4.0(0.5)   & $ -0.5(0.2) $ &   9b  \\
     -  &  6545      &   4.00       & $ -0.35     $ &   10  \\  \hline
 55057  &  7580(250) &   3.6(0.5)   & $ +0.14(0.21)$&   1a  \\  \hline
 57006  &  6180(70)  &   3.60(0.15) & $ -0.08(0.11)$&   1a  \\ 
     -  &  6100(100) &   3.50(0.15) & $ -0.04(0.11)$&   1b  \\  \hline 
171834  &  6840(200) &   4.6(0.5)   & $ -0.25(0.21)$&   1a  \\
     -  &  6670      &   4.05       & $ -0.42     $ &    2  \\
     -  &  6700(150) &   3.9(0.25)  & $ -0.5(0.15)$ &   9a  \\
     -  &  6750(250) &              & $           $ &   9b  \\  \hline
184663  &  6600(200) &   4.5(0.5)   & $ -0.19(0.21)$&   1a  \\
     -  &  6535      &   4.22       & $ +0.03     $ &    2  \\  \hline
\hline

\end{tabular}
\end{table}

\begin{table}
\caption{The references for Table~\ref{tab:study}.\label{tab:study2}}
\begin{tabular}{r|lllr}
\multicolumn{5}{l}{} \\ 
\multicolumn{5}{l}{Ref.\ 1a: Optimal parameters found using method A (\vwa)} \\
\multicolumn{5}{l}{Ref.\ 1b: Fixed parameters used for method B} \\
\multicolumn{5}{l}{Ref.\ 1c: Optimal parameters found using method C}\\
\multicolumn{5}{l}{Ref.\  2: Balachandran (1990)} \\ 
\multicolumn{5}{l}{Ref.\  3: Bruntt \etal\ (2002)} \\
\multicolumn{5}{l}{Ref.\  4: Edvardsson \etal\ (1993)} \\
\multicolumn{5}{l}{Ref.\  5: Favata \etal\ (1997)} \\
\multicolumn{5}{l}{Ref.\  6: Fisher, D.\ ({private communication})} \\
\multicolumn{5}{l}{Ref.\  7: Gratton \etal\ (1996)} \\
\multicolumn{5}{l}{Ref.\  8: Hartmann \& Gehren (1988)} \\ 
\multicolumn{5}{l}{Ref.\ 9a: Lastennet \etal\ (2001) (colours)} \\
\multicolumn{5}{l}{Ref.\ 9b: Lastennet \etal\ (2001) (spectroscopy)}\\
\multicolumn{5}{l}{Ref.\ 10: Perrin (1976)} \\ 
\multicolumn{5}{l}{Ref.\ 11: Vogt \etal\ (2002)} \\
\multicolumn{5}{l}{Ref.\ 12: Zakhozhaj \& Shaparenko (1996)} 
\end{tabular}
\end{table}

\subsection{Previous spectroscopic studies}

Lastennet~et~al.~(2001) (hereafter \last) have analyzed some of the potential
\corot\ targets and they used the same spectra we have used in this study:
we have six stars in common with \last\ (compared in Table~\ref{tab:study}).
\last\ have also used the \halpha\ hydrogen lines to determine \teff.
The fact that the \halpha\ line covers three Echelle orders makes the
determination of the continuum a difficult task, hence
the error on \teff\ is around 250~K.
\last\ have compared Johnson
and Str\"omgren photometric indices of
observations and calculated model atmospheres in order to
determine \teff, \logg, and \feh. 
The \teff\ estimates made by 
\last\ agree well with the results from \templogg\ although the
two different techniques (\halpha\ lines and photometry) show
systematic differences at the 100~K level.
\last\ have used Fe\,\iet\ and Fe\,\ii\ lines in a limited spectral region
to estimate \feh\ and \logg\ to within 0.2~dex and 0.5~dex and these results
also agree with \templogg\ within the error bars. 
The exception is for their star HD~46304 but this is probably due 
to its high \vsini~$=200$~\kms.

Vogt \etal\ (2002) have found a low-mass companion around
HD~43587 in a campaign to search for exo-planets.
From their Keck spectra they find
\teff\ $= 5795$~K, [Fe/H] $= -0.03$, and 
\vsini\ $= 2.7$ \kms\ (Vogt \etal\ 2002 have not 
quoted errors on \teff\ and metallicity).
D. Fisher (private communication) has re-analysed the 
Keck spectrum and finds \teff $= 5931\pm100$ and [Fe/H] $= 0.00\pm0.05$.

The extensive catalogue by Cayrel de Strobel \etal\ (2001)
contains fundamental atmospheric parameters found in the literature
based on analyses of medium-high resolution spectra. 
The catalogue contains results from several
studies for most of the stars examined here, but
unfortunately, no individual error estimates are 
given by the original authors.
In Table~\ref{tab:study} we summarize the parameters from these
studies (Ref.~2, 4, 5, 7 \& 8) along with our new results (Ref.\ 1a--c).
The references in column five of Table~\ref{tab:study} are 
listed in detail in Table~\ref{tab:study2}.
Note that for \last\ we give two results:
(9a) which corresponds to their result when comparing observed and
synthetic colour-indices in $B-V$, $U-B$, and $b-y$; (9b) \teff\ is 
determined from the \halpha\ line while \logg\ and [Fe/H] 
are determined from a small sample of Fe\,\iet\ and
Fe\,\ito\ lines in a region around 6130~\AA.

\section{Abundance analysis\label{sec:methods}}

Nine stars were analysed with the semi-automatic
software \vwa\ (H.~Bruntt -- HB) while two groups 
(I.~F.~Bikmaev -- IFB -- and M.~Gillon/P.~Magain -- MG/PM) 
made an independent analysis by means of using 
the equivalent width of isolated lines for 
some of the slowly rotating stars. 
We will describe each method here before comparing the results.
Firstly, we shall discuss the atmospheric models we have applied.

\subsection{Atmospheric models\label{models}}

For method A and B (see below) we have used a modified version of 
the \atlasnine\ code (Kurucz~1993) for the calculation of the
atmospheric models (Kupka 1996, Smalley \& Kupka 1997) in which
turbulent convection theory in the formulation of 
Canuto \& Mazzitelli (1991, 1992) is implemented.
For method C the standard \atlasnine\ models (Kurucz 1993)
with mixing length parameter $\alpha=1.25$ were used.


For the extraction of models in a grid with different
\teff\ and \logg\ we have 
used interpolation in the grid of models 
(also modified \atlasnine\ models) distributed by Heiter \etal\ (2002).
We have used these interpolated models to make abundance analysis 
in order to constrain the fundamental parameters in 
\sect~\ref{constrain}.

\subsection{Automatic abundance analysis with \vwa\ (Method~A)\label{vwa_desc}}

We have used the semi-automatic
software package \vwa\ (Bruntt \etal~2002) to carry out the 
abundance analysis of nine \corot\ targets.
For each star the software selects the least blended lines 
from atomic line list
data extracted from the \vald\ data base (Kupka \etal\ 1999). 
The atomic data consist of the element name and ionization state, 
wavelength, excitation potential, oscillator strength, and damping parameters.
For each selected line the synthetic spectrum is calculated 
and the input abundance is changed iteratively until the equivalent 
width of the observed and synthetic line match. We used {\sc synth} 
(version $2.5$, see Valenti \& Piskunov~1996) to calculate 
the synthetic spectrum. 
This software was kindly provided by N.~Piskunov (private communication).

\subsubsection{Rotational velocities}

In order to be able to compare the observed and 
synthetic spectra the latter is convolved by the instrumental
profile (approximated by a Gaussian) and 
a rotational broadening profile. 
The projected rotational velocities 
were determined by fitting the synthetic
spectrum of a few of the least blended lines 
to the observed spectrum by convolving with 
different rotation profiles. 
Note that we have used zero macroturbulence, and thus
our quoted values for \vsini\ is a combination of the effects of
rotational broadening and macroturbulence.
The accuracy of \vsini\ by this method is about 5--10\%.
\last\ used a more refined method 
(see Donati \etal\ 1997) but
our results agree within the estimated errors.
The exception is for HD~43587 for which we have found 7 \kms\ 
while Vogt \etal\ (2002) find \vsini\ $=2.7$ \kms.
Note that D.\ Fisher finds \vsini\ $= 2.2$~\kms\ from the same spectrum
(private communication). These lower values agrees with \last\ and 
consequently we have used a low value of \vsini\,$=2.5$~\kms.
The reason for the apparent discrepancy is simply 
the limit of the spectral resolution of the \elodie\ spectrograph 
\ie\ $v = c / R \simeq 7$~\kms. Note that since \vwa\ relies on the 
measurement of equivalent widths small errors in \vsini\ will 
have a negligible effect on the derived abundances.

\begin{figure}
\hskip 0.3cm
\includegraphics[width=8.8cm]{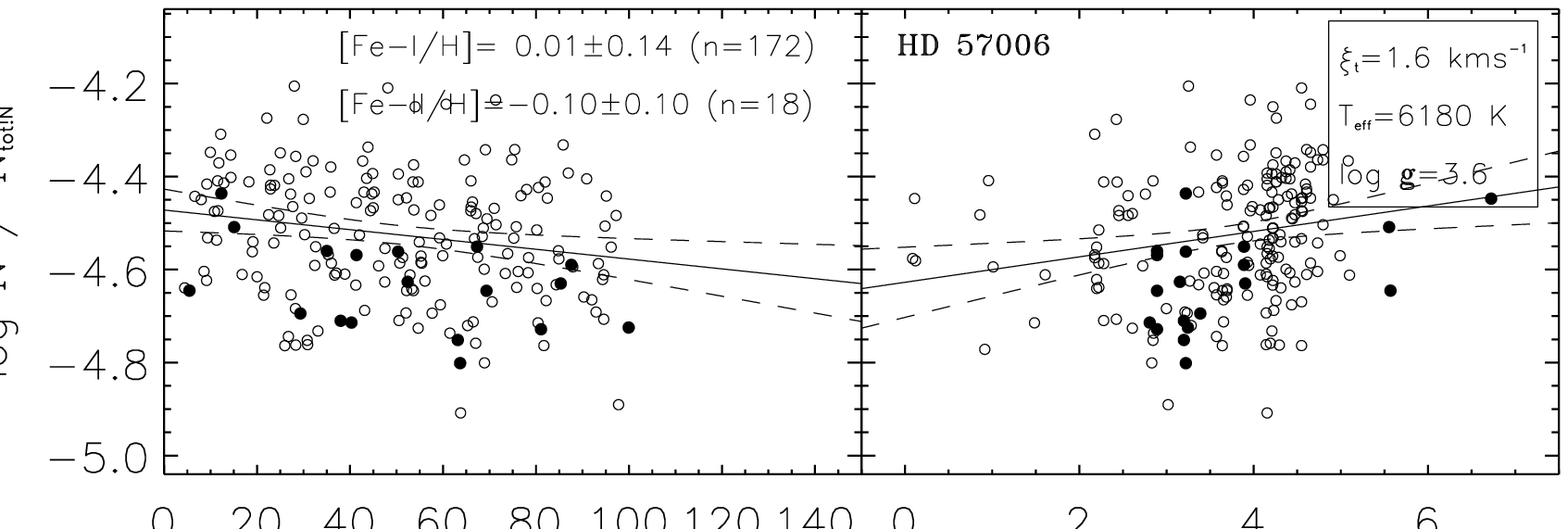} 
\vskip 0.5cm
\hskip 0.3cm
\includegraphics[width=8.8cm]{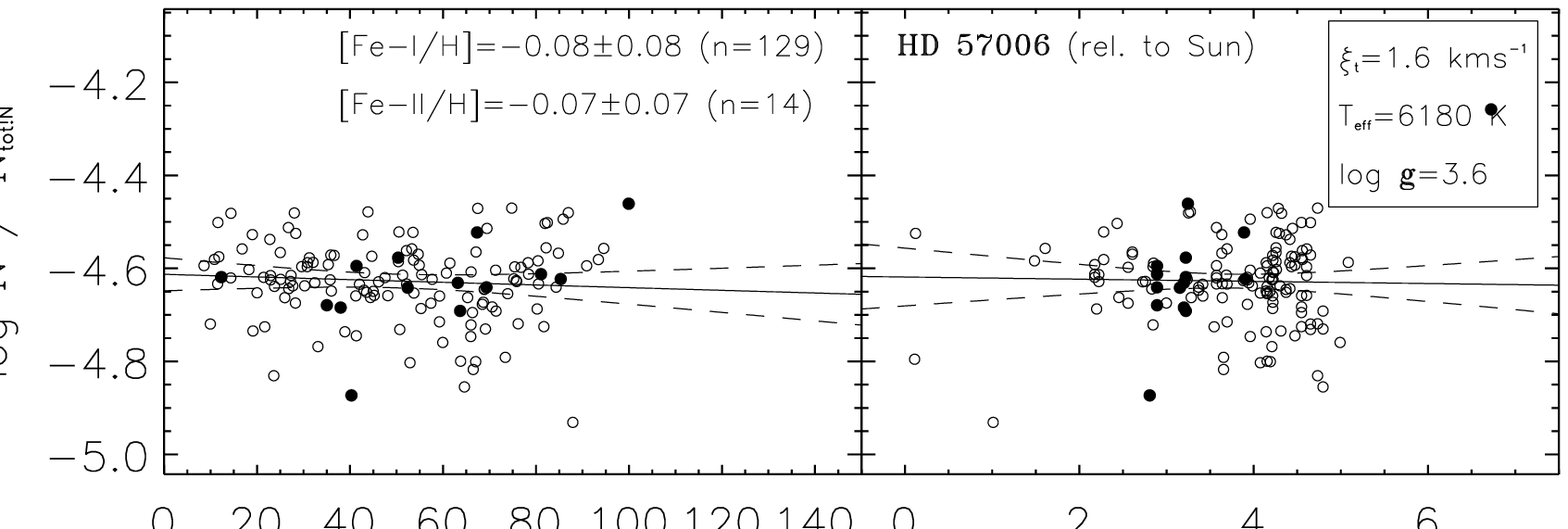} 
\vskip 0.2cm
    \caption{Iron abundances from lines in the spectrum of HD~57006. 
Open symbols are neutral iron lines and solid symbols are {Fe {\sc ii}} lines. 
The \topp\ panel shows the raw abundances 
while in the \bott\ panel the abundances are relative
to the abundances (line-by-line) found for the Sun.
The abundances are plotted versus
equivalent width and excitation potential. 
The mean abundances and standard deviation are given in each panel and the
atmospheric model parameters are found in the boxes.
The straight lines are linear fits to the neutral iron lines and the 
dashed curves indicate the 3\,$\sigma$ interval for the fit.
\label{fig:hd57006}}
\end{figure}

\subsubsection{Measuring abundances relative to the Sun\label{sec:relsun}}

During our analyses with \vwa\ we have found that when 
measuring abundances relative to the same
lines observed in the Sun there is a dramatic decrease in the 
error. Since the abundance of the Sun is 
well-known (\eg\ Grevesse \& Sauval 1998) we have decided to make
a differential analysis.
There are several sources of error that may affect the
derived abundances but the most important error is 
erroneous \loggf\ values. Incorrect removal
of scattered light when reducing the spectra will cause systematic
errors. 

In Fig.~\ref{fig:hd57006} we show the iron abundance 
for several lines versus both equivalent width and lower excitation 
potential for HD~57006. 
In the \topp\ panel we show the direct results while in the \bott\ panel
the abundances are measured relative to the results for the Sun (line-by-line).
For the zero point we used $\log N_{\rm Fe}/N_{\rm tot}=-4.54$ from 
Grevesse \& Sauval~(1998).
The open symbols are used for neutral ion lines (Fe\,{\sc i}) and
filled symbols for ionized Fe lines (Fe\,{\sc ii}).

Note the dramatic $\sim$40\% decrease in the uncertainty for the 
neutral ion lines when measuring relative to the Sun (\bott\ panel
in Fig.~\ref{fig:hd57006}). Also, the discrepancy in abundance 
found from Fe\,{\sc i} and Fe\,{\sc ii} lines is removed.
For the applied atmospheric model of HD~57006 there is no correlation 
with either equivalent width nor excitation potential. Also, both Fe\,{\sc i} and 
Fe\,{\sc ii} give the same result which indicates that the
applied model is correct.

\begin{table*}
 \centering
    \caption{Comparison of the abundances of the main elements found by three
different methods. For each star and element we give the 
difference in abundance $\Delta A$ in the sense 
method A {\em minus} method B or C.
The numbers in parenthesis are the combined \rms\ standard deviations,
but only given if at least five lines are used for both methods.
For example, $\Delta A = -0.03(14)$ for {V~{\sc i}} in column HD~43587$^{A-C}$ means
the difference in abundance found with method A and C is $-0.03\pm0.14$.
The number $n$ is the number of lines used for each method.
This data is plotted in Fig.~\ref{fig:bikmaev_bruntt}.
 \label{tab:bikmaev_bruntt}}
 \setlength{\tabcolsep}{3pt} 
 \begin{footnotesize}
\begin{tabular}{l|lr|lr|lr|lr|lr}
           & \multicolumn{2}{c|}{HD 43587$^{\rm A-C}$} & \multicolumn{2}{c|}{HD 43587$^{\rm A-B}$} & \multicolumn{2}{c|}{HD 45067$^{\rm A-B}$} & \multicolumn{2}{c|}{HD 57006$^{\rm A-B}$} & \multicolumn{2}{c}{     Sun$^{\rm A-B}$}  \\
\hline
           & $\Delta A$ & $n_A/n_C$ &  $\Delta A$ & $n_A/n_B$ & $\Delta A$ & $n_A/n_B$ & $\Delta A$ & $n_A/n_B$ & $\Delta A$ & $n_A/n_B$ \\
\hline
  {C  \sc  i} & $-0.08    $ &     3/2 & $+0.13    $ &     3/2 & $+0.29    $ &     3/3 & $+0.37    $ &     3/2 & $+0.27    $ &     3/2  \\ 
  {Na \sc  i} & $+0.01    $ &     4/2 & $+0.04    $ &     4/3 & $+0.04    $ &     4/3 & $+0.14    $ &     4/2 & $+0.02    $ &     4/2  \\ 
  {Mg \sc  i} &         $-$ &     $-$ & $+0.21    $ &     1/1 & $+0.04    $ &     1/2 & $+0.08    $ &     1/1 & $+0.18    $ &     2/2  \\ 
  {Al \sc  i} & $-0.01    $ &     2/2 & $-0.01    $ &     2/2 & $-0.30    $ &     2/2 & $-0.29    $ &     1/2 & $-0.24    $ &     3/2  \\ 
  {Si \sc  i} & $+0.03(7)$  &    29/5 & $-0.06(19)$ &   29/25 & $-0.22(19)$ &   25/16 & $-0.12(14)$ &   22/16 & $-0.12(13)$ &   33/20  \\ 
  {Si \sc ii} &         $-$ &     $-$ & $-0.25    $ &     3/2 & $-0.14    $ &     3/2 & $-0.15    $ &     2/2 & $-0.11    $ &     3/2  \\ 
  {S  \sc  i} &         $-$ &     $-$ & $+0.04    $ &     1/4 & $-0.07    $ &     1/4 & $+0.08    $ &     1/3 & $-0.01    $ &     1/5  \\ 
  {Ca \sc  i} & $+0.03   $  &    11/4 & $+0.05(12)$ &    11/8 & $-0.15(15)$ &     8/8 & $-0.00(14)$ &    11/7 & $-0.07(18)$ &    12/9  \\ 
  {Sc \sc ii} & $+0.06(5)$  &     7/5 & $-0.10(13)$ &    7/12 & $-0.18(13)$ &    5/12 & $+0.02(20)$ &    7/10 & $-0.06(17)$ &    7/14  \\ 
  {Ti \sc  i} & $+0.00(13)$ &    37/9 & $+0.04(14)$ &   37/55 & $-0.06(15)$ &   26/42 & $+0.10(13)$ &   12/29 & $+0.03(13)$ &   45/81  \\ 
  {Ti \sc ii} & $+0.01    $ &    11/1 & $-0.06(17)$ &   11/24 & $-0.13(20)$ &    8/24 & $+0.04(9)$  &    7/16 & $-0.02(9)$  &   14/28  \\ 
  {V  \sc  i} & $-0.03(14)$ &     8/7 & $+0.08(15)$ &    8/18 & $-0.06(11)$ &    6/14 & $+0.07    $ &     4/7 & $+0.01(11)$ &   10/27  \\ 
  {Cr \sc  i} & $+0.00(9)$  &   20/12 & $+0.01(10)$ &   20/43 & $-0.15(17)$ &   15/44 & $+0.02(16)$ &   15/21 & $-0.03(15)$ &   22/68  \\ 
  {Cr \sc ii} & $-0.03(14)$ &     7/6 & $-0.17(16)$ &     7/9 & $-0.26(18)$ &    7/14 & $-0.10(12)$ &     5/9 & $-0.07(7)$  &    7/15  \\ 
  {Mn \sc  i} & $-0.12   $  &     7/4 & $-0.16(16)$ &    7/20 & $-0.29    $ &    4/22 & $-0.20    $ &    4/14 & $-0.16(19)$ &    7/26  \\ 
  {Fe \sc  i} & $+0.03(7)$  &  206/57 & $-0.08(17)$ & 206/213 & $-0.19(18)$ & 178/244 & $-0.09(16)$ & 147/162 & $-0.09(15)$ & 209/335  \\ 
  {Fe \sc ii} & $+0.05(8)$  &    23/6 & $-0.04(16)$ &   23/78 & $-0.08(14)$ &   21/35 & $+0.02(14)$ &   17/26 & $-0.02(11)$ &   26/41  \\ 
  {Co \sc  i} & $-0.15(5)$  &     8/5 & $-0.11(16)$ &    8/35 & $-0.14(29)$ &    7/26 & $+0.18    $ &    3/12 & $-0.01(17)$ &   12/47  \\ 
  {Ni \sc  i} & $-0.01(9)$  &   61/31 & $-0.08(12)$ &   61/85 & $-0.15(14)$ &   45/78 & $-0.03(14)$ &   34/44 & $-0.05(11)$ &   67/97  \\ 
  {Cu \sc  i} & $-0.08    $ &     2/1 & $-0.14    $ &     2/3 & $-0.33    $ &     2/3 & $+0.10    $ &     2/2 & $+0.09    $ &     3/2  \\ 
  {Zn \sc  i} &         $-$ &     $-$ & $-0.08    $ &     2/3 & $-0.22    $ &     2/3 & $-0.20    $ &     2/3 & $-0.10    $ &     2/3  \\ 
  {Y  \sc ii} & $+0.05    $ &     5/2 & $-0.13(17)$ &    5/11 & $-0.15(18)$ &    6/12 & $+0.02(20)$ &     6/8 & $-0.13(18)$ &    6/14  \\ 
  {Ba \sc ii} &         $-$ &     $-$ & $+0.01    $ &     1/3 & $-0.33    $ &     1/2 & $-0.19    $ &     1/1 & $-0.21    $ &     3/2  \\ 
\end{tabular}
\end{footnotesize}
\end{table*}

For the results presented in this work we have measured 
abundances relative to the abundance found for the same lines 
in the spectrum of the Sun.
However, we have not done this for the star HD~49933 and 
the stars with higher \vsini\ since we see no significant 
improvement for these stars. 
The reason is that these stars are much hotter than the 
reference star ($\Delta T_{\rm eff} > 1000$~K). 
This means that the Sun and the hotter stars have 
relatively few lines in common which are suitable 
for abundance analysis.
However, it is likely that this is an indication 
of inadequacies of the model atmospheres. 



\subsubsection{Adjusting the microturbulence}

For each model we adjust the microturbulence $\xi_t$
until we see no correlation between the abundances
and equivalent widths found for 
the Fe\,\iet, Cr\,\iet, and Ni\,\iet\ lines. 
To minimize the effect of saturated lines
we only use lines with equivalent 
width below 100~m\AA\ and for Fe we only use lines with 
excitation potential in the range 2--5~eV to minimize 
the effect of erroneous \teff\ of the model atmosphere.
For some stars only Fe could be used due to 
a lack of non-blended Cr and Ni lines.
For the slowly rotating stars ($\vsini<25$~\kms) the 
error on $\xi_t$ is about $0.1-0.2$~\kms, for the
stars rotating moderately fast ($50<$ \vsini\ $<85$~\kms)
the error is $0.5$~\kms, and $0.7$~\kms\ for HD~55057 which has
$\vsini=120$~\kms.  The contribution to the error on the
abundances from the uncertainty of the microturbulence is
about 0.03 dex for Fe for the slow rotators
while for stars with the highest \vsini\ (HD~49434 and HD~55057)
the errors in the abundances are of the order 0.07--0.10 dex.

For each star we made the abundance analysis for a grid 
of models in order to be able to constrain \teff\ and 
\logg\ which will be discussed in \sect~\ref{constrain}.






\subsection{"Classical" abundance analysis (Method B)}

"Classical" abundance analysis was applied (IFB) to
three of the slowly rotating 
\corot\ targets -- HD~43587, HD~45067, and HD~57006 -- and 
to the solar spectrum (observed reflection from the Moon).

Model atmospheres were calculated as described in 
\sect~\ref{models} with
the adopted parameters of \teff, \logg, and solar composition.
Effective temperatures were determined by using Johnson and 
\str\ color-indices extracted from \simbad\ data base and 
the \logg\ parameters were determined by using \hipparcos\ parallaxes 
as outlined in \sect~\ref{hipp}.
The values we have used are given in Table~\ref{tab:study}
marked with Ref.~1b. These parameters were not changed
during our analysis with method B.

Equivalent widths of all identified unblended and some partially blended
spectral lines were measured by using {\sc dech} software 
based on PC (Galazutdinov 1992).
Abundance calculations were performed in the LTE approximation by using
\widthni\ (Kurucz 1993; with modifications by V.~Tsymbal 
and L.~Mashonkina for PC, private communication).
Atomic parameters of the spectral lines were extracted from the \vald\ data base
with corrections of oscillator strengths in a few cases as described
in the paper of Bikmaev \etal\ (2002).

Lines with equivalent widths $<$~100~m\AA\ were used when possible to decrease
influences of microturbulence and inaccurate damping constants.
For each star the microturbulence was chosen as to minimize
the correlation between the abundances and equivalent
widths found for Fe, Cr, Ti, Ni, and Co lines.

\subsection{Abundance analysis after 
            continuum re-normalisation (Method C)\label{sec:magain}}

A somewhat different approach was followed by two of us (MG and PM)
and tested on HD~43587.

First, the continuum was redetermined on the basis of a number of 
pseudo-continuum windows selected from inspection of the 
Jungfraujoch solar atlas (Delbouille \etal\ 1973).  These
windows were selected to be as close as possible to the true continuum 
and the mean level was measured in each of them.  
In all cases, it is between 99\% and 100\% of the true
continuum.  The same windows are used for the program star, after 
correction for its radial velocity and after checking that no 
telluric lines enter the window because of
the Doppler shift.  The mean flux is then measured in the stellar 
windows and a table is constructed, containing the ratio of 
the mean flux in the star versus the mean flux in
the Sun. A spline curve is then fitted through these points and the 
stellar spectrum is divided by that curve, thus providing our 
re-normalized spectrum.  It may differ from the
normalization described in Sect.\  2.1 by $\sim 1\%$, which is not 
negligible at all when weak lines are considered.

Then, a list of very good lines is selected by inspection of the solar 
atlas, and equivalent widths (EWs) are measured by least-squares 
fitting of Gaussian and Voigt profiles.  Voigt profiles are 
generally necessary to adequately fit the lines in the
very high resolution solar atlas, as well as for the medium and strong 
lines in the stellar spectrum.  For the weakest stellar lines, 
Gaussians are generally adequate.
The fitted profile is compared to the observed one and the line is 
rejected if any significant discrepancy appears (\eg\ line asymmetries).

Both the star and the Sun are analysed by using models extracted from 
the same grid (\atlasnine\ models; Kurucz 1993).  
The line oscillator strengths are adjusted so that the solar 
EWs are reproduced, using the adopted solar model
and the known solar atmospheric parameters and abundances
(similar to what is described in Sect.~\ref{sec:relsun}).
Whenever possible, we use damping constants as calculated by 
Anstee \& O'Mara (1995), which have proved to be 
quite accurate (Anstee \etal\ 1997 and Barklem \& O'Mara 2000).

The microturbulence is determined in order to remove 
any correlation of the computed abundance with the line EW, 
using sets of lines from the same ion and
similar excitation potentials.  We chose to determine the effective 
temperature by pure spectroscopic means, using excitation equilibria, 
\ie\ by ensuring that abundances deduced from lines of the 
same ionic species do not depend on the line excitation
potential.  The surface gravity was determined from ionization 
equilibria, by ensuring that lines of neutral and ionic 
species of the same element give the same abundance.


\begin{figure}
\hskip 0.3cm
\includegraphics[width=8.4cm]{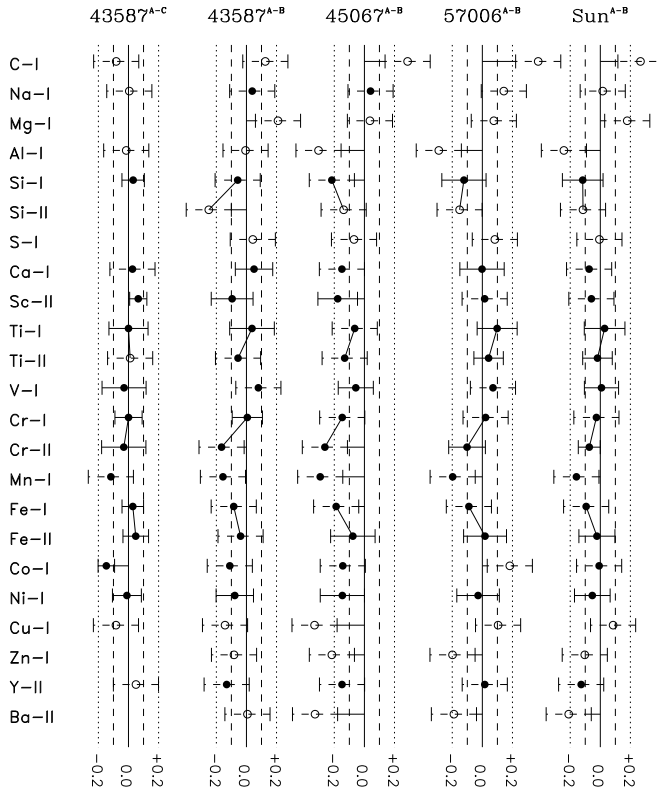} 
    \caption{Comparison of the abundance analysis results for the three
different methods. We show the differences in the overall abundances 
for several elements in the sense method A {\em minus} method B or C
(as indicated above the star name).
For HD~43587 we have compared the result from method A (\vwa) with both
method B and C, while for the remaining stars we only compare method A and B.
Open symbols indicate that less than five lines were used. The error
bars are the combined standard deviation of the mean for the two methods.
\label{fig:bikmaev_bruntt}}  
\end{figure}

As an illustration of the coherence of the method, let us consider the 
determination of the most critical parameter, 
\ie\  $T_{\rm eff}$.  We used three excitation
equilibria, which give the following $T_{\rm eff}$: $5866 \pm 18$~K (Fe\,\iet), 
$5850 \pm 102$~K (Cr\,\iet) and $5880 \pm 57$~K (Ni\,\iet).  
The error bars are deduced from the uncertainty in
the slope of a straight line fitted to 
abundances {\em vs}.\ excitation potential.  
A weighted mean gives $T_{\rm eff} = 5867\pm17$\,K, taking 
into account the three individual error bars as well as the scatter 
in the three determinations.  The internal 
error bar on $T_{\rm eff}$, taking into account the uncertainties in the other 
parameters, amounts to 20\,K.  The other parameters 
are: $\log{g} = 4.29 \pm 0.06$ and $v_{\rm turb} = 2.13 \pm 0.1$ km/s.  



\subsection{Comparison of the abundance analysis methods\label{sec_comp_meth}}

The three methods described above have different 
advantages and disadvantages.
Most importantly, \vwa\ (method A) relies on the computation of the 
synthetic spectrum 
and thus this method can be used for stars 
with moderately high \vsini\ when a mild degree of blending of lines is tolerated.
Also, \vwa\ is a semi-automatic program and the user may easily
inspect if the fitted lines actually match the observed spectrum.
When comparing the observed and synthetic spectrum 
obvious problems with the continuum level can be found: 
errors of just 2\,\% of the continuum level will give 
large differences in the derived abundance --- 
perhaps as much as 0.1--0.2~dex and such lines are rejected
after visual inspection.

Method B and C both rely on measuring equivalent widths. 
It is a much faster method computationally, but special care 
has to be taken to avoid systematic errors from line blends.

In Fig.~\ref{fig:bikmaev_bruntt} and in Table~\ref{tab:bikmaev_bruntt} 
we show the comparison of the
final abundance analysis results using the three methods described above. 
We show the differences in abundance in the sense ``method A'' {\em minus} 
``method B''. We show results for four different stars with 
low \vsini\ for which both these methods are applicable.
For the star HD~43587 we also show result ``method A'' {\em minus} 
``method C''.
The differences refer to the mean abundances, \ie~the lines are not
necessarily the same in the different analyses. 

The plotted error bars in Fig.~\ref{fig:bikmaev_bruntt} are
calculated as the quadratic sum of the standard deviation 
of the mean for each method. In the cases
where we have fewer than five lines we have no good estimate of the
error, but from the elements with more lines we find that an overall
error of 0.15 dex seems plausible. Thus, in these cases
we have plotted a dashed error bar corresponding to 0.15 dex.


The comparison of method A and C for HD~43587 shows excellent
agreement for all elements despite the quite different approaches.
We emphasize that \teff\ and \logg\ have been 
adjusted as a part of the analysis for these two methods and
the agreement is remarkable, 
$\Delta$\teff$ = 3\pm62$~K, $\Delta$\logg$=-0.09\pm0.16$, 
$\Delta$[Fe/H]$ = -0.04\pm0.14$; see Table~\ref{tab:study}
for the individual parameters (Ref.~1a and 1c).
However, the results from method C have significantly lower
formal error on the fundamental parameters. It is likely
that the renormalization of the spectrum done with
method C is the reason for this improvement, and should be
considered in future analyses. The good agreement in the
derived fundamental parameters may be an indication that the
errors quoted in Table~\ref{tab:study} for method~A (ref.~1a) 
are too large.

In this context we mention Edvardsson et al. (1993) who 
carried out an abundance study of 189 slowly rotating F and G type 
stars using data of similar quality.
Edvardsson et al. (1993) used fixed values for the atmospheric 
parameters and found abundances from a small number of lines. 
Their quoted {\em rms} scatter on the abundances is lower than we find here, but
this is may be due to the small number of carefully selected lines.
For example Edvardsson et al. (1993) use typically $\sim15$ Fe\,\iet\ lines 
while in this study we use $\sim150-200$ Fe\,\iet\ lines.
More importantly, Edvardsson et al. (1993) normalized their spectra by defining
``continuum windows'' from spectra of the Sun and Procyon. The fact that
Edvardsson et al. (1993) obtain abundances with smaller scatter may indeed be
due to their careful normalization of the spectra.



The comparison between method A and B show significant 
systematic differences of the order $0.05-0.10$ dex between 
neutral and ionized abundances of Ti, Cr, and Fe. 
The worst case is for star HD~45067 where we find a 
systematic offset of 0.1~dex between method A and B. 
The reason for these discrepancies is that the models 
used by \vwa\ and the classical 
analysis have slightly different parameters, \ie\ differences in
\logg\ and \teff\ of the order 0.1-0.2 dex and 80-100~K.

Based on the generally good agreement between the three analyses, we
have used the results from method A (\vwa) for all the stars 
since they were analysed in a homogeneous way. 
In \sects~\ref{constrain} and \ref{sec:abundances} we 
will describe in detail the analysis of the results from \vwa.


\begin{figure}
\hskip 0.2cm
\includegraphics[width=8.1cm]{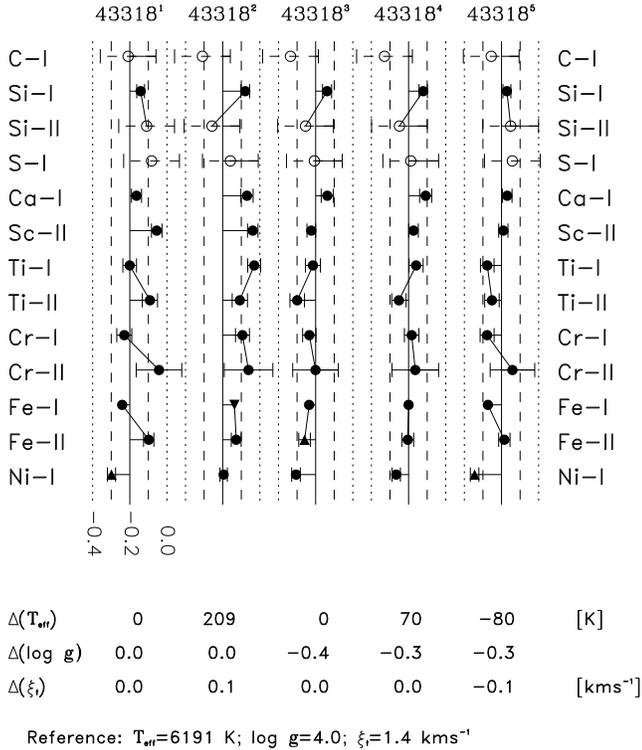}
    \caption{Abundances for selected elements for five different models
of HD~43318. 
Open symbols are used when less than five lines were used. Circle
symbols are used when no significant correlation of abundance and 
lower excitation potential is found. The triangle symbol 
pointing down (up) are used when a significant negative (positive) 
correlation is found (see \eg\ Fe {\sc i} in model 2) 
The atmospheric parameters of the
reference model is given below each panel and the relative
parameters ({\em model - reference}) is given below the abundances.
\label{fig:models43318}}  %
\end{figure}

\begin{figure}
\hskip 0.7cm
\includegraphics[width=7.2cm]{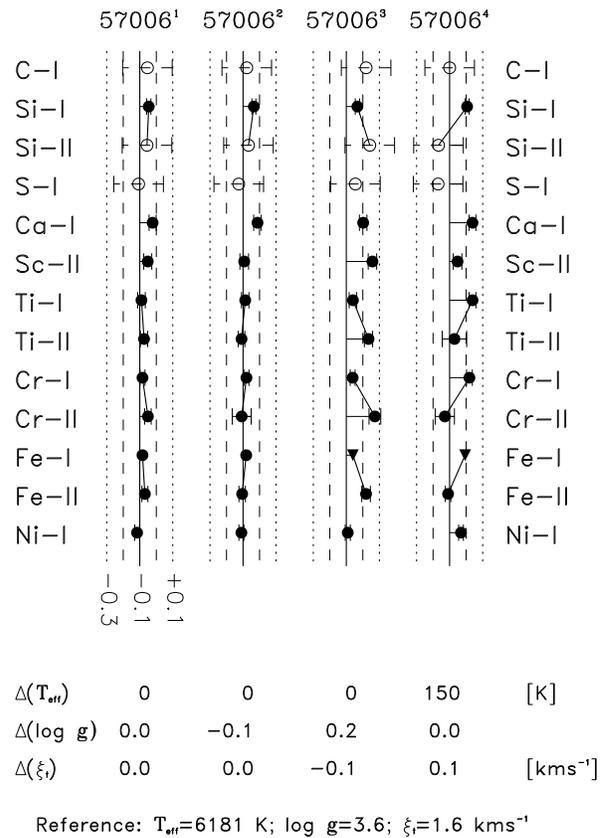}
    \caption{Abundances for selected elements for four different models
for HD~57006. See caption of Fig.~\ref{fig:models43318} for details.
\label{fig:models57006}}  
\end{figure}


\section{Constraining model parameters\label{constrain}}

In order to measure the sensitivity of the derived abundances on
the model parameters we 
have made the abundance analysis for a grid of models for each star.
For this purpose we used the \vwa\ software.
For the initial model we have used \logg\ values from \hipp\ parallaxes and
metallicity from \templogg. For \teff\ we used results from \templogg\ for 
the hot stars but we used the result from line depth ratios
for the cooler stars (as discussed in Sect.~\ref{sec:fundamental}).
We have then carried out abundance analysis of
models with lower and higher values of \teff\ and \logg, \ie\ at
least five models for each star.
For each model we adjust the microturbulence to minimize the
correlation between abundance of 
Fe\,\iet, Cr\,\iet, and Ti\,\iet\ lines and the measured 
equivalent width. 



Changes in the model temperature will affect the depth of neutral lines while
changes in \logg\ will mostly affect the lines of the ionized elements.
Furthermore, changes in \teff\ will affect the correlation 
of Fe abundances (and in some cases Cr and Ni)
and the lower excitation potential ($E_{\rm low}$; \cf\ the Boltzmann equation).
By requiring that there must be no correlation with \exita\
and the least possible difference between neutral and ionized 
lines of the same element (from this point we call this ``ion-balance'')
we have constrained \teff\ and \logg. 
In the next Sections we will discuss the results for the stars with 
slow (\vsini\,$<25$\,\kms) and moderate rotation.


\subsection{Constraining \teff\ and \logg\ for the slow rotators\label{sec:slowrot}}

To give an impression of how well we can constrain the fundamental
atmospheric parameters and the uncertainty of the derived abundances,
we show the abundances found for different models for the stars 
HD~43318 in Fig.~\ref{fig:models43318}
and HD~57006 in Fig.~\ref{fig:models57006}
As discussed in Sec.~\ref{sec:relsun} the abundance of
each line is measured relative to the abundance found from
the spectrum of the Sun where the zero points are 
from Grevesse \& Sauval~(1998). Open symbols indicate
that less than five lines are used for the abundance estimate.
Triangles are used to indicate a significant ($> 1\,\sigma$)
slope when fitting a line to abundance \vs\ \exita.
The triangle symbol points down (up) if the correlation 
is negative (positive). 

We will discuss in detail the results for HD~43318 shown
in Fig.~\ref{fig:models43318} for five different models. 
The parameters of the reference model is given below the plot
($T_{\rm eff}=6191$\,K, $\log g=4.0$, and microturbulence $\xi_t=1.4$\,\kms)
and the adjustments for the different models are given as
$\Delta\teff$, $\Delta\logg$, and $\Delta\xi_t$. 
For example, in model 2, we have added $209$\,K to \teff. 
This essentially creates ion-balance for Fe and Cr but not for Ti.
But a higher \teff\ results in a 
negative correlation for Fe\,\iet\ abundance and \exita\ which 
is marked by a triangle symbol pointing down in Fig.~\ref{fig:models43318}. 
For model 3 we have decreased \logg\ which also 
restores the ion-balance, but
now a positive correlation of Fe\,\ito\ and \exita\ is the result. 
The model with $\Delta T_{\rm eff}=70$\,~K and $\Delta\log g = -0.3$ dex 
is our preferred model. 

The results of four models of HD~57006 are compared in
Fig.~\ref{fig:models57006}. These models have a much 
better agreement in the ion balance for both 
Si, Ti, Cr, and Fe.  
It is seen that changing \logg\ by 0.2~dex 
or \teff\ by 150~K in the model clearly breaks the
ion balance and also results in correlations with
Fe\,\iet\ abundance and \exita.

\begin{figure*}
\centering
\vskip -0.7cm
\includegraphics[width=13.5cm]{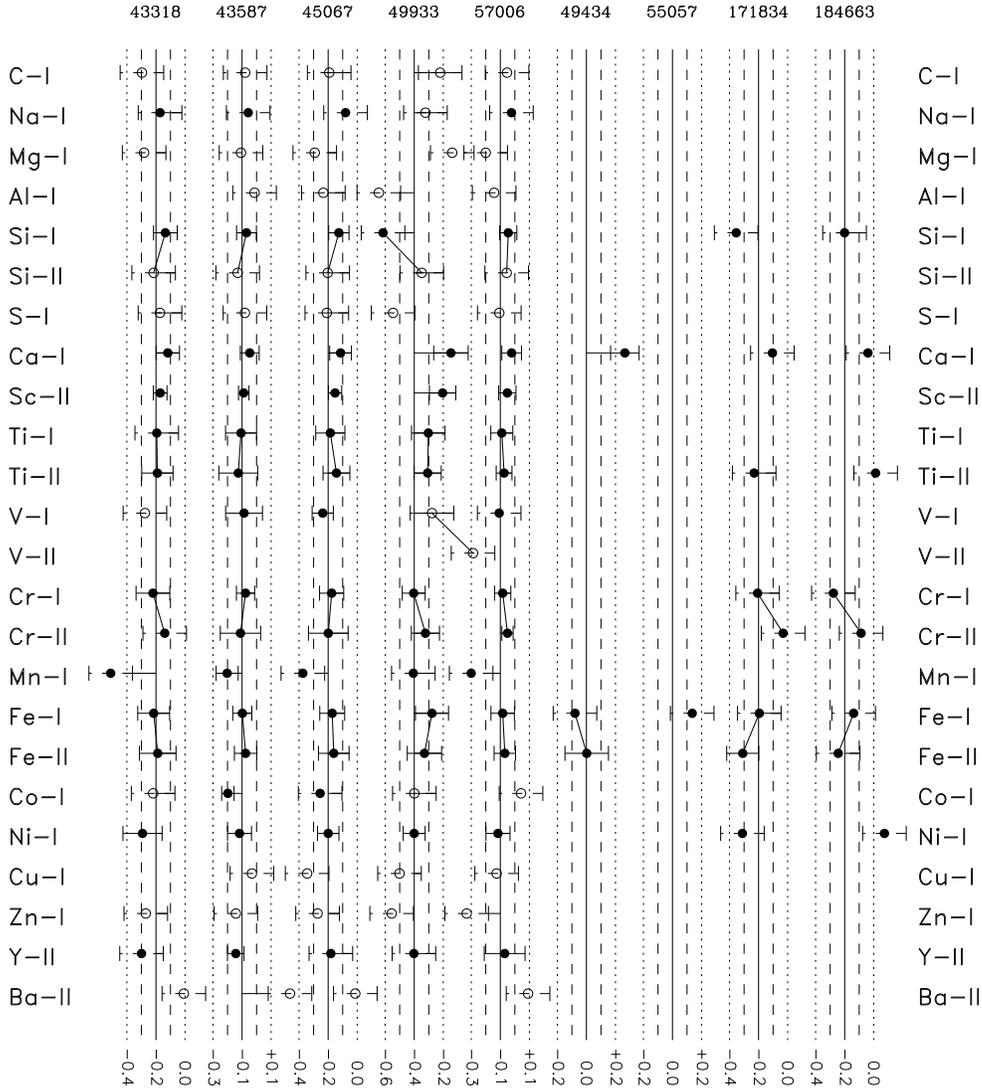} 
    \caption{Results of the abundance analysis of the proposed
\corot\ main targets. For each star we plot the derived abundance relative to the
Sun (Grevesse \& Sauval~1998). The open symbols indicate that three or less lines were used.
The error bars correspond to the \rms\ errors but only if the error is smaller than 0.15 dex and 
more than five lines were used. Otherwise a dashed error bar corresponding to 0.15 dex is plotted. 
Elements of the same species are connected with a solid line.
The results shown here are also given in Table~\ref{tab:vwa_results}). 
\label{fig:vwa_results}}  
\end{figure*}

For the remaining slowly rotating stars
we show the abundance pattern for the 
best models in the column 1--5 in Fig.~\ref{fig:vwa_results}. 
It can be seen that the ion balance for Fe, Cr, and Fe 
is within the errors bars in all cases for these stars.


\subsection{Constraining \teff\ and \logg\ for moderate rotators\label{sec:fastrot}}

Column 6--9 in Fig.~\ref{fig:vwa_results} shows the 
abundance pattern for the four stars with moderate \vsini.


The star HD~49434 is a moderately fast rotator ($v \sin i = 84$\,\kms).
Here we find no usable Cr lines and thus have only used 
the ion balance for Fe to adjust \teff\ and \logg, but with
the constraint that there must be no correlation of Fe\,\iet\ 
abundance and \exita. This approach, however, 
is connected with large uncertainties due to the blending of lines.
HD~49434 was also analysed by Bruntt \etal\ (2002). 
In the present study we have included more Fe\,\iet\ lines 
but our result agrees well with Bruntt \etal\ (2002).

The star HD~55057 has $v \sin i=120$~\kms\ and there are only 
very few lines usable for abundance analysis.

For the moderately high rotators HD~171834 and HD~184663 
($v \sin i \simeq 50-60$\,\kms) there
is a highly significant difference in the abundance found
from neutral and ionized lines of Cr and Fe. Adjusting \logg\
and \teff\ cannot produce a coherent result. The \logg\
we have found for HD~171834 is 0.5~dex higher than what we
find from the \hipp\ parallax and also results from 
the literature (\cf\ Table~\ref{tab:study}),
but it is still within the large error of 0.5~dex.

\subsection{Summary of the parameter estimation}

In Sect.~\ref{sec:slowrot} we have estimated how well
we can constrain \teff\ and \logg\ for two slow rotators.
From a similar analysis for all five slow rotators ($v \sin i < 15$~\kms)
we find that we can constrain \teff\ and \logg\ to within $70-100$~K and 
$0.1-0.2$~dex, respectively.
However, these error estimates
do not include possible systematic errors due to the 
simplifications of the applied 1-D LTE atmospheric models. 
We are planning a future study of the importance of the applied models.

We have four stars in our sample 
with moderately high \vsini\ which
makes it difficult to constrain the atmospheric parameters.
The errors on \teff, \logg, and [Fe/H]
are about $200$~K, $0.5$~dex, and 0.15~dex, respectively.

With the analyses provided here \logg\ cannot be 
constrained better than with the parallax 
from \hipparcos\ but it gives us an 
independent estimate. On the other hand
we determine \teff\ more accurately than from
\str\ photometry but not better than the temperatures
from the line depth ratios.
From the many iron lines the metallicity we find is
determined to within 0.1~dex for the slow rotators 
(\vsini\ $< 15$~\kms) and 0.2~dex for the faster rotators.
In the former case this is a factor of two better than the
metallicity determined from \str\ photometry.

Our final \teff\ and \logg\ estimates are given in Table~\ref{tab:study}
under Ref.~1a along with the [Fe/H] found from the abundance 
analysis (\cf\ \sect~\ref{sec:abundances}).


\begin{table*}
\centering
\caption{Results of the abundance analysis of \corot\ targets using \vwa\ (method A).
For all stars except HD~49933, HD49434, HD55057, HD171834, and HD184663 the abundances
are measured relative to the Sun, \ie\ the mean of the line-by-line differences. For the
other stars the abundances are given relative to the standard solar 
abundances given in first column (Grevesse \& Sauval 1998).
The \rms\ errors are found in the parentheses and the number of 
lines used, $n$, is given. For example, 
the result ``$-0.12(8)$ $8$'' for Ca in HD~43318 means
$\log N_{\rm Ca} / N_{\rm tot} - (\log N_{\rm Ca} / N_{\rm tot})_\odot = -0.12\pm0.08$
as found from eight lines.
\label{tab:vwa_results}}
\setlength{\tabcolsep}{3pt} 
\begin{footnotesize}
\begin{tabular}{c|l|lr|lr|lr|lr|lr|}

                             &                &
         \multicolumn{2}{c|}{HD 43318} & \multicolumn{2}{c|}{HD 43587} & 
         \multicolumn{2}{c|}{HD 45067} & \multicolumn{2}{c|}{HD 49933} & \multicolumn{2}{c|}{HD 57006} \\
$\log (N/N_{\rm tot})_\odot$ &  Element       & 
         {$\Delta$A} & $n$ &  {$\Delta$A} & $n$ & {$\Delta$A} & $n$ & {$\Delta$A} & $n$ & {$\Delta$A} & $n$ \\
\hline
$ -3.49$& {C  \sc  i} & $-0.30    $ &       2 & $-0.08(8)$  &       3 & $-0.19(16)$ &       3 & $-0.22    $ &       2 & $-0.05(9)$  &       3   \\ 
$ -5.71$& {Na \sc  i} & $-0.17(17)$ &       4 & $-0.06(4)$  &       4 & $-0.08(12)$ &       4 & $-0.32    $ &       1 & $-0.02(13)$ &       4   \\ 
$ -4.46$& {Mg \sc  i} & $-0.28    $ &       2 & $-0.11    $ &       1 & $-0.29    $ &       1 & $-0.14    $ &       2 & $-0.20    $ &       1   \\ 
$ -5.57$& {Al \sc  i} &         $-$ &     $-$ & $-0.01    $ &       2 & $-0.23    $ &       2 & $-0.64    $ &       1 & $-0.14    $ &       1   \\ 
$ -4.49$& {Si \sc  i} & $-0.14(8)$  &      15 & $-0.07(6)$  &      29 & $-0.13(7)$  &      25 & $-0.61(36)$ &      20 & $-0.05(5)$  &      22   \\ 
$ -4.49$& {Si \sc ii} & $-0.22    $ &       2 & $-0.13(9)$  &       3 & $-0.20    $ &       3 & $-0.35(8)$  &       3 & $-0.06    $ &       2   \\ 
$ -4.71$& {S  \sc  i} & $-0.17    $ &       1 & $-0.08    $ &       1 & $-0.21    $ &       1 & $-0.55    $ &       1 & $-0.11    $ &       1   \\ 
$ -5.68$& {Ca \sc  i} & $-0.12(8)$  &       8 & $-0.05(6)$  &      11 & $-0.11(7)$  &       8 & $-0.15(11)$ &      18 & $-0.02(6)$  &      11   \\ 
$ -8.87$& {Sc \sc ii} & $-0.17(4)$  &       5 & $-0.09(3)$  &       7 & $-0.15(4)$  &       5 & $-0.20(9)$  &       7 & $-0.05(5)$  &       7   \\ 
$ -7.02$& {Ti \sc  i} & $-0.19(16)$ &      19 & $-0.11(10)$ &      37 & $-0.19(10)$ &      26 & $-0.30(11)$ &      26 & $-0.09(7)$  &      12   \\ 
$ -7.02$& {Ti \sc ii} & $-0.19(10)$ &      10 & $-0.13(13)$ &      11 & $-0.14(9)$  &       8 & $-0.31(9)$  &      23 & $-0.07(5)$  &       7   \\ 
$ -8.04$& {V  \sc  i} & $-0.28    $ &       3 & $-0.09(12)$ &       8 & $-0.24(7)$  &       6 & $-0.28    $ &       1 & $-0.11(12)$ &       4   \\ 
$ -8.04$& {V  \sc ii} &         $-$ &     $-$ &         $-$ &     $-$ &         $-$ &     $-$ & $+0.00    $ &       1 &         $-$ &     $-$   \\ 
$ -6.37$& {Cr \sc  i} & $-0.22(11)$ &      12 & $-0.08(6)$  &      20 & $-0.18(8)$  &      15 & $-0.40(7)$  &      20 & $-0.08(5)$  &      15   \\ 
$ -6.37$& {Cr \sc ii} & $-0.14(26)$ &       5 & $-0.11(13)$ &       7 & $-0.20(13)$ &       7 & $-0.32(9)$  &      19 & $-0.05(3)$  &       5   \\ 
$ -6.65$& {Mn \sc  i} & $-0.51(14)$ &       4 & $-0.20(7)$  &       7 & $-0.37(9)$  &       4 & $-0.41(17)$ &      10 & $-0.30(2)$  &       4   \\ 
$ -4.54$& {Fe \sc  i} & $-0.22(10)$ &     141 & $-0.10(6)$  &     206 & $-0.17(8)$  &     178 & $-0.28(11)$ &     218 & $-0.08(8)$  &     147   \\ 
$ -4.54$& {Fe \sc ii} & $-0.19(12)$ &      18 & $-0.08(7)$  &      23 & $-0.16(10)$ &      21 & $-0.33(11)$ &      34 & $-0.07(7)$  &      17   \\ 
$ -7.12$& {Co \sc  i} & $-0.22(25)$ &       3 & $-0.20(4)$  &       8 & $-0.26(22)$ &       7 & $-0.40    $ &       2 & $+0.04(16)$ &       3   \\ 
$ -5.79$& {Ni \sc  i} & $-0.29(13)$ &      36 & $-0.12(8)$  &      61 & $-0.20(7)$  &      45 & $-0.40(7)$  &      35 & $-0.12(8)$  &      34   \\ 
$ -7.83$& {Cu \sc  i} &         $-$ &     $-$ & $-0.03    $ &       2 & $-0.34    $ &       2 & $-0.50    $ &       1 & $-0.13    $ &       2   \\ 
$ -7.44$& {Zn \sc  i} & $-0.27    $ &       2 & $-0.14    $ &       2 & $-0.27    $ &       2 & $-0.56(6)$  &       3 & $-0.33    $ &       2   \\ 
$ -9.80$& {Y  \sc ii} & $-0.30(18)$ &       4 & $-0.14(5)$  &       5 & $-0.18(16)$ &       6 & $-0.40(8)$  &       4 & $-0.07(14)$ &       6   \\ 
$ -9.91$& {Ba \sc ii} & $-0.01    $ &       1 & $+0.23    $ &       1 & $-0.01    $ &       1 & $+0.48(23)$ &       3 & $+0.09    $ &       1   \\ 


\end{tabular}

\vskip 0.3cm
\hskip 1.3cm

\begin{tabular}{c|l|lr|lr|lr|lr|}

      &     &  \multicolumn{2}{c|}{HD 49434} & \multicolumn{2}{c|}{HD 55057} & \multicolumn{2}{c|}{HD 171834} & \multicolumn{2}{c|}{HD 184663}  \\ 

$\log (N/N_{\rm tot})_\odot$ &  Element       & 
         {$\Delta$A} & $n$ &  {$\Delta$A} & $n$ & {$\Delta$A} & $n$ & {$\Delta$A} & $n$  \\

\hline
$-4.49$& {Si \sc  i}  &         $-$ &     $-$ &         $-$ &     $-$ & $-0.35(68)$ &       9 & $-0.20(60)$ &      10   \\ 
$-5.68$& {Ca \sc  i}  & $+0.26(9)$  &       8 &         $-$ &     $-$ & $-0.11(40)$ &      10 & $-0.04(23)$ &       9   \\ 
$-7.02$& {Ti \sc ii}  &         $-$ &     $-$ &         $-$ &     $-$ & $-0.23(24)$ &      13 & $+0.01(29)$ &       9   \\ 
$-6.37$& {Cr \sc  i}  &         $-$ &     $-$ &         $-$ &     $-$ & $-0.21(60)$ &       6 & $-0.28(28)$ &       7   \\ 
$-6.37$& {Cr \sc ii}  &         $-$ &     $-$ &         $-$ &     $-$ & $-0.03(24)$ &       9 & $-0.09(40)$ &       6   \\ 
$-4.54$& {Fe \sc  i}  & $-0.08(22)$ &      37 & $+0.14(29)$ &      19 & $-0.20(30)$ &      60 & $-0.14(31)$ &     102   \\ 
$-4.54$& {Fe \sc ii}  & $+0.00(14)$ &       7 &         $-$ &     $-$ & $-0.31(11)$ &       7 & $-0.24(27)$ &       9   \\ 
$-5.79$& {Ni \sc  i}  &         $-$ &     $-$ &         $-$ &     $-$ & $-0.31(43)$ &      18 & $+0.07(40)$ &      12   \\ 

\end{tabular}
\end{footnotesize}
\end{table*}

\section{Abundances of the \corot\ targets\label{sec:abundances}}

In \sect~\ref{sec:methods} three different methods to compute
abundances were applied to a subset of our sample, and we
obtained similar results within the error bars. 
For the analysis of the whole sample 
method A (\vwa\ semi-automatic software) has been 
adopted and the results are given here.

Abundances relative to the Sun for nine 
potential \corot\ targets is 
shown in Fig.~\ref{fig:vwa_results}.
The first five stars plotted (HD~43318 to HD~57006) are the
slow rotators for which the best results are obtained.
From the abundance patterns we find no evidence 
of classical chemically peculiar stars.

For the slowly rotating stars, the lines of Mn 
give a systematically lower abundance, indicating a
systematic error. We have found that for this specific element, 
the hyperfine structure levels are not always found 
in \vald, thus yielding an erroneous result. We also
find a relative over-abundance of Ba, but we only have
few lines for this element.

The abundances are also 
given in Table~\ref{tab:vwa_results}. The parameters of the
atmospheric models used are given in Table~\ref{tab:study} under
Ref.~1a. The abundances are given relative to the 
solar abundances found by Grevesse \& Sauval~(1998) which
are found in the first column in Table~\ref{tab:vwa_results}.
In the parentheses we give the \rms\ standard 
deviation based on the line-to-line
scatter, but only if at least four lines were used.
As an example we find for Ca\,\iet\ in HD~43318 the result $-0.12(8)$ which means
$\log N_{\rm Ca} / N_{\rm tot} - (\log N_{\rm Ca} / N_{\rm tot})_\odot = -0.12\pm0.08$.
We give the \rms\ error from 8 lines, and the standard deviation of the mean is 0.03 dex.
The number of lines used to determine the abundance is in the column with label $n$.

In addition to the quoted errors contributions from the
abundance zero points (Grevesse \& Sauval~1998) of the order
0.05~dex and model dependent uncertainties of the 
order 0.05-0.10 (0.20)~dex 
for the slow (fast) rotators must be added. 
Thus, we can constrain the metallicity [Fe/H] to 
about 0.10 (0.20)~dex. In Table~\ref{tab:study} we have 
given [Fe/H] as the mean of the iron abundance relative
to the Sun found from Fe\,\iet\ and Fe\,\ito\ lines 
for all three methods described in Sect.~\ref{sec:methods}.

\section{Conclusions and future prospects\label{conclusions}}

\begin{itemize}
\item  We have performed a detailed abundance analysis of nine potential
   \corot\ main targets. The accuracy of the abundances of the 
   main elements is of the order 0.10~dex when including possible 
   errors on microturbulence and inadequacies of the applied 1D LTE 
   atmospheric models. 

\item We have compared three different methods for the analysis 
   which show very good agreement. The discrepancies are due to
   the different parameters we have used for the stellar
   atmosphere models.

\item We have found no evidence for chemically peculiar stars.

\item We have constrained \teff, \logg, and metallicity 
      to within $70-100$~K, $0.1-0.2$~dex, and 0.1~dex for
      the five slow rotators in our sample. 
      For the four moderate rotators we cannot constrain the 
      fundamental parameters very well, \ie\ \teff, \logg, and [Fe/H]
      to within $200$~K, $0.5$~dex, and 0.15~dex.

\item For most of the stars 
      our results for the fundamental parameters agree 
      with the initial estimates from Str\"omgren photometry, 
      line depth ratios, and \halpha\ lines, and the 
      \hipp\ parallaxes. For HD~43318 we
      have found a somewhat lower \teff\ and \logg. 
      For HD~49933 we find a \teff\ about 200~K hotter 
      than previous studies.
      For the fast rotators HD~171834 and HD~184663
      we find a quite high \logg\ value compared to other methods,
      but the uncertainty on our estimate is large (0.5 dex).

\item  For the two \corot\ targets HD~46304 and HD~174866 abundance 
       analyses were not possible due to the very high \vsini.

\end{itemize}

Suggestions for future studies of the \corot\ targets:

\begin{itemize}

\item From the comparison of two independent
analyses (method A and C; \cf~\ref{sec_comp_meth}) 
we have found evidence that a careful (re-)normalization of
the spectra may be very important but this must be investigated further.

 \item To probe the interior of the stars with the 
    asteroseismic data from \corot\ we need to know the 
    abundances of elements which affect the evolution 
    of the stars, namely C, N, and O.
    Thus, spectra that cover the infrared regions with 
    good C, N, and O lines must be obtained.


 \item Recent 3D atmospheric models should also be used in 
       the analysis to explore the importance of the models.



\end{itemize}

We finally note that the next paper in this series
will give the results of the abundance analysis for some
recently proposed \corot\ primary targets,  
the possible secondary \corot\ targets, as well as 
the proposed targets for the \romer\ mission.


\begin{acknowledgements}

IFB and HB are grateful to Tanya Ryabchikova for her
advice on various stages of abundance determination.
We are grateful to Nikolai Piskunov for providing us with his
software for the calculation of synthetic spectra ({\sc synth}). 
Thanks to V.~V.~Kovtyukh for supplying his line depth ratio calibrations.
HB is supported by NASA Grant NAG5-9318,
IFB is supported by RFBR grant 02-02-17174, and IFB and WWW were
supported by grant P-14984 of the 
Fonds zur F\"orderung der wissenschaftlichen Forschung.
This research has made use of the SIMBAD database, operated at CDS, Strasbourg, France.

\end{acknowledgements}


\end{document}